\documentclass[manuscript]{aastex}
\usepackage[utf8]{inputenc}
\usepackage{graphicx}
\usepackage{amsmath} 
\usepackage{wasysym} 
\usepackage{bm}  

\newcommand{\up}[1]{\ifmmode^{\rm #1}\else$^{\rm #1}$\fi}

\newcommand{\arcd}{\ifmmode^{\circ}\else$^{\circ}$\fi}
\newcommand{\arcm}{\ifmmode{'}\else$'$\fi}
\newcommand{\arcs}{\ifmmode{''}\else$''$\fi}

\slugcomment{\today{} version}

\shorttitle{OGLE-2008-BLG-092 -- triple microlens}
\shortauthors{Poleski et al.}

\begin{document}


\title{Triple Microlens OGLE-2008-BLG-092L: Binary Stellar System with a Circumprimary Uranus-type Planet}

\author{Radosław Poleski\altaffilmark{1,2}, Jan Skowron\altaffilmark{2}, Andrzej Udalski\altaffilmark{2}, Cheongho Han\altaffilmark{3},} 
\author{Szymon Koz\l{}owski\altaffilmark{2}, \L{}ukasz Wyrzykowski\altaffilmark{2,4}, Subo Dong\altaffilmark{5,6}, Michał K. Szyma\'nski\altaffilmark{2},} 
\author{Marcin Kubiak\altaffilmark{2}, Grzegorz Pietrzyński\altaffilmark{2,7}, Igor Soszyński\altaffilmark{2}, Krzysztof Ulaczyk\altaffilmark{2},}
\author{Pawe\l{} Pietrukowicz\altaffilmark{2}, and Andrew Gould\altaffilmark{1}}
\email{poleski@astronomy.ohio-state.edu}

\altaffiltext{1}{Department of Astronomy, Ohio State University, 140 W. 18th Ave., Columbus, OH 43210, USA}
\altaffiltext{2}{Warsaw University Observatory, Al. Ujazdowskie 4, 00-478 Warszawa, Poland}
\altaffiltext{3}{Department of Physics, Chungbuk National University, Cheongju 371-763, Korea}
\altaffiltext{4}{Institute of Astronomy, University of Cambridge, Madingley Road, Cambridge CB3 0HA, UK}
\altaffiltext{5}{Kavli Institute for Astronomy and Astrophysics, Peking University, Yi He Yuan Road 5, Hai Dian District, Beijing 100871, China}
\altaffiltext{6}{Institute for Advanced Study, Einstein Drive, Princeton, NJ 08540, USA}
\altaffiltext{7}{Universidad de Concepci\'on, Departamento de Astronomia, Casilla 160–C, Concepci\'on, Chile}


\begin{abstract}
We present the gravitational microlensing discovery of 
a $4~{\rm M_{\rm Uranus}}$ planet that  
orbits a $0.7~{\rm M_{\odot}}$ star at $\approx18~{\rm AU}$. 
This is the first known analog of Uranus. 
Similar planets, i.e., cold ice giants, are inaccessible to either radial velocity or transit methods because of the long orbital periods, 
while low reflected light prevents direct imaging. 
We discuss how similar planets may contaminate the sample of the very short microlensing events that are interpreted as free-floating planets with an estimated rate of $1.8$ per main sequence star. 
Moreover, the host star has a nearby stellar (or brown dwarf) companion. 
The projected separation of the planet is only $\approx 3$ times smaller than that of
the companion star, suggesting significant dynamical interactions.
\end{abstract}


\section{Introduction} 

The solar system planets are divided into three groups: 
small rocky planets, massive gas giants, and intermediate mass ice giants. 
These groups differ not only in mass but also in location of their orbits. 
The orbits of all rocky planets lie within the snow-line i.e., a boundary beyond which ices can condense resulting in much more efficient planetary formation. 
The snow-line lies at $2.7~{\rm AU}$ in the solar system and increases with stellar mass \citep{kennedy08}. 
Beyond the snow-line all planets have masses larger than the limit for runaway accretion which is $\approx 10{\rm M_{\oplus}}$. 
Gas giants have the highest masses and their orbits are only a small factor larger than the snow-line distance ($1.9$ for Jupiter and $3.5$ for Saturn). 
By contrast, the ice giants have smaller masses, but still above $10{\rm M_{\oplus}}$, and lie significantly further away: Uranus is seven times farther than the snow-line distance and Neptune is 11 times farther. 

In contrast to other solar system planets, ice giants could not have formed close to their present location. 
The lifetime and the density of a protoplanetary disk were not large enough for
planetesimals to accrete sufficient mass for in situ formation \citep{pollack96}. 
The models that assume the formation of Uranus and
Neptune close to Jupiter and Saturn not only predict currently observed orbital separations, but
are also consistent with the observed properties of the Kuiper belt \citep{thommes99,thommes02}, and require that Jupiter
and Saturn have crossed the 2:1 mean motion resonance \citep{tsiganis05}. 
The solid-to-gas ratio puts additional constraints on Uranus and Neptune formation \citep{goldreich04,helled14}.

The process of planetary formation cannot be well understood without studies of planets similar to those in the solar system. 
The {\it Kepler} satellite has revealed extrasolar analogs of rocky planets \citep{borucki12,quintana14}. 
Objects with masses comparable to Jupiter and Saturn have been discovered using radial velocities, and some of these have
orbits larger than the snow-line of the parent stars \citep{butler06}. 
Gravitational microlensing also detects analogs of Jupiter and Saturn \citep[e.g., ][]{bond04,gaudi08}. 
However, both transit and radial velocity methods fail when analogs of Uranus and Neptune are searched for \citep{kane11}, and no such object is known. 
The orbital periods of these planets are longer than the human lifetime. 
Thus, methods that depend on periodic phenomena like the change in the host radial velocity or a transit in front of the star would require extremely long experiments in order to find such planets. 
On the other hand, direct imaging is capable of discovering planets that are far away from the hosts and does not require data taken over whole orbital period for clear planet detection. 
Nonetheless, the planets that are  directly observed are much more massive and hotter 
objects that inhabit young systems and thus are different from Uranus and Neptune. 
Thus, the only method that can discover analogs of Uranus and Neptune is gravitational microlensing \citep[see review by][]{gaudi12}.  
Instead of looking for periodic phenomena, microlensing uses light from a fortuitously aligned background source to obtain a snapshot of a planetary system's gravitational potential. 
This allows planets to be discovered irrespective of their orbital period. 

Here, we report the microlensing discovery of the first planet that has mass and orbital separation similar to Uranus, 
and thus likely shares similar origin. 
The microlensing event  OGLE-2008-BLG-092 showed a separate lensing signal from each of the lens components. 
The large time difference between the planet and host microlensing subevents indicates that the projected separation is about five times larger than the Einstein radius $\theta_{\rm E}$, 
which sets the angular scale of the event. 
For typical disk lenses the Einstein radius corresponds to $r_{\rm E} = D_l\theta_{\rm E} \approx 3~{\rm AU}$, where $D_l$ is the distance to the lens. 
This means that the planetary orbit is very wide.
We confirm the large value of projected separation by detailed modeling of the event. 
This raises a question of planet formation. 
A third microlensing subevent observed on the same source revealed the presence of the stellar companion to the planet host. 

In the next section we present observations of the event. 
Very simple yet complete microlensing and physical parameters are presented in Section~3.1 in a form that is accessible for broad readership. 
More detailed analysis is presented in Sections~3.2 and 4. 
In Section 5, 
we discuss planetary properties and formation, as well as   
the prospects for finding similar planets in the future, 
compare the planetary anomaly in this event to free-floating planets, 
and remark on the long-term stability of the system.
We conclude in Section 6.

\section{Observations} 

\begin{figure}
\includegraphics[bb=45 17 574 743,angle=270,width=.95\textwidth]{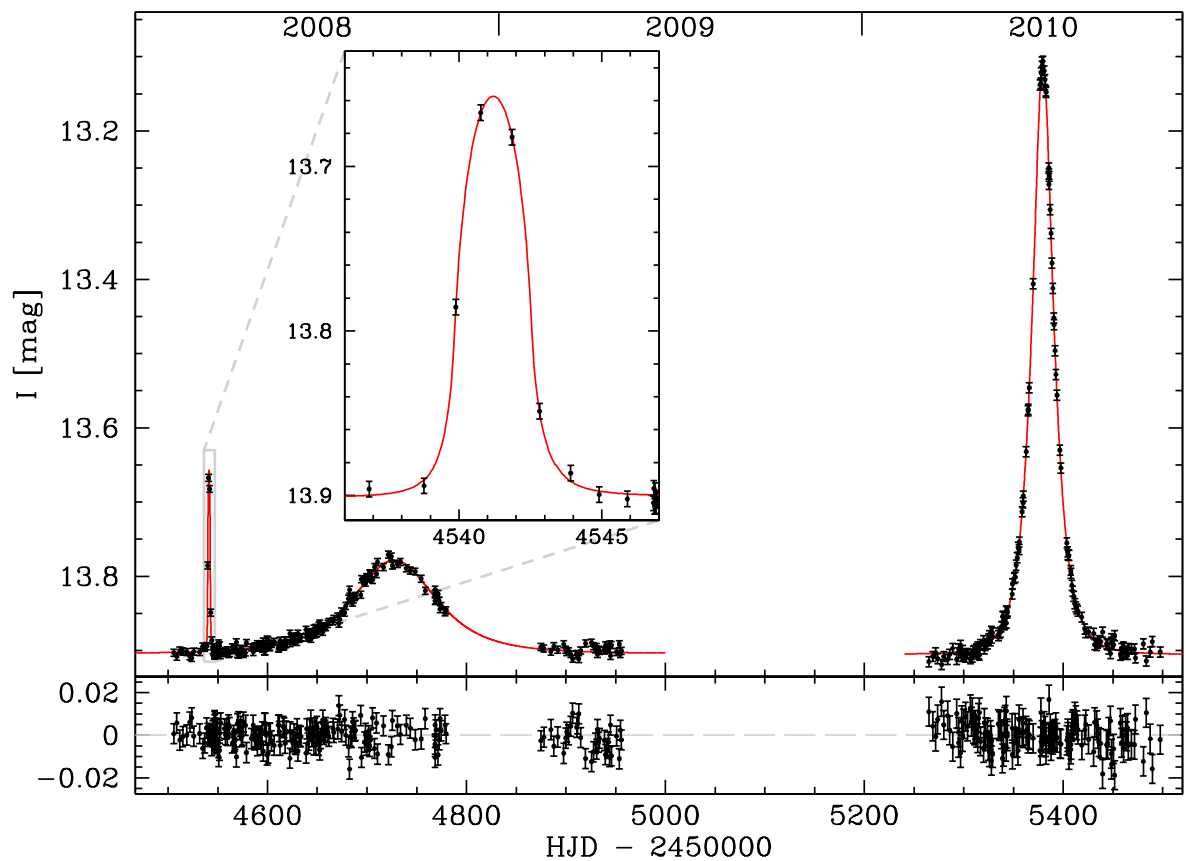}
\caption{Light curve of OGLE-2008-BLG-092 microlensing event. 
Models fitted to OGLE-III (Section~3.2) and OGLE-IV (Section~3.1) photometry are presented by lines. 
The bottom panel shows the models residuals. 
The inset shows the planetary subevent. 
The OGLE-IV baseline flux was aligned to the OGLE-III for plotting. 
\label{fig:lc}}
\end{figure}

The Optical Gravitational Lensing Experiment (OGLE) discovered a microlensing 
event on ${\rm HJD'} \equiv {\rm HJD} - 2450000 = 4542.103$ at 
RA: $17^{\rm h}47^{\rm m}29\fs42$, Dec. $-34\degr43\arcmin35\farcs6$ ($l\approx-4\fdg75$, $b\approx-3\fdg34$) 
based on a $\approx 2$ day long planetary anomaly. 
Although only four epochs were collected during this time, 
the anomaly led to the announcement of the event OGLE-2008-BLG-092 by the OGLE Early Warning System \citep{udalski03}.  
Following this initial bump, 
a standard microlensing was observed with relatively low magnification and timescale slightly longer  than is typical, see Figure~\ref{fig:lc}. 
At this point, the possibility that the lens could be composed of a star with a distant planet was recognized, but could not be verified based on the existing data. 

The data were collected by the third phase of OGLE project (OGLE-III) 
that used the $1.3~{\rm m}$ Warsaw Telescope situated in the Las Campanas Observatory, Chile.
The telescope was equipped with an eight chip CCD camera. 
The pixel scale was $0\farcs26$ and the total field of view was $36' \times 36'$. 
For details of the instrumentation setup, see \citet{udalski03}. 
In 2009, OGLE-III ended and the camera was replaced by a four times larger instrument, and the fourth phase of the survey (OGLE-IV) started. 
The new camera has the same pixel size of $15~{\rm \mu m}$ and the same pixel scale. 
Its 32 $4{\rm k} \times 2{\rm k}$ pixel CCD chips give a total field of view of $1.4~{\rm deg}^2$. 
As before, the $I$ band filter is used for most observations. 
The photometry was performed using Difference Image Analysis \citep{alard98,alard00,wozniak00}.
The observations resumed in 2010. 
The same year one more microlensing event at the position of OGLE-2008-BLG-092 was observed.  
Its presence was not recognized at that time because OGLE online data reduction and microlensing alerts started a year later. 
The last subevent had a larger amplitude than the previous one. 
It was found that the event was not significantly blended, which turns out to be crucial in ruling  out multi-source models, as we show below. 
This  established clearly that the 2008 subevents were caused by a lens composed of a star with a distant planet. 
The 2010 subevent revealed a companion to the 2008 lens that is either a low mass star or a brown dwarf. 
Thanks to the weakness of the interactions of the three masses composing the lens system, all important properties of the lens system can be derived based on simple light curve inspection. 
This is in contrast to previously analyzed triple lens microlensing events, i.e., 
OGLE-2006-BLG-109 \citep{gaudi08,bennett10},
OGLE-2012-BLG-0026 \citep{han13}, and
OGLE-2013-BLG-0341 \citep{gould14a}. 
The configuration of the lens in this system is one of the simplest that is possible for a triple lenses i.e., each lens component produces a separate microlensing subevent. 

OGLE-2008-BLG-092 was discovered on a relatively bright star of $I=13.9~{\rm mag}$. 
Very few microlensing events have been observed on such bright stars \citep{wyrzykowski14_tmpA}. 
The source star was also observed in the first two phases of the OGLE project \citep{udalski92,udalski97}. 
The pixel scales of 
those observations were $0\farcs44$ and $0\farcs417$, respectively. 
The reference images constructed using the best-quality frames have seeing full width at half maximum of $1\farcs12$ and $1\farcs14$, respectively, and correspond to $16$ and $11$ years before the event, respectively. 
Inspection of these reference images does not reveal elongated profile of the source star that would constrain the models if the lens were a nearby fast-moving object (see Section~\ref{sec:det}).
The OGLE-II time-series photometry 
did not reveal any brightness changes. 
In the OGLE-I data the source star is slightly overexposed. 

This analysis is based on the 2007-2009 OGLE-III data and the first two seasons of OGLE-IV data. 
The online OGLE-III photometry that facilitated the real-time detection of microlensing events was re-reduced to correct for a low-level systematic noise. 
The 2009 data end in May when the OGLE-III camera was dismounted. 
The errorbars for all photometric points were set to the dispersion of the baseline data (OGLE-III: $4.7~{\rm mmag}$, OGLE-IV: $6.8~{\rm mmag}$). 
This is justified  because the event was not blended and not highly magnified. 
The final dataset consisted of 354 OGLE-III and 383 OGLE-IV photometric points. 
The OGLE-III data were transformed to the standard system as described by \citet{udalski08red} and \citet{szymanski11}. 
The zero point of the OGLE-IV photometry was transformed in a similar manner. 
No significant dependence of the correction on the $(V-I)$ color was found, while the dispersion of the brightness differences of nearby stars was $0.013~{\rm mag}$. 
The baseline brightness of this event is $0.02~{\rm mag}$ brighter than in the OGLE-III i.e., $1.7\sigma$. 

\section{Microlensing model} 

\subsection{Basic model} 

In strong contrast to all previous 
triple and the majority of binary microlensing events,
the essential physics of this event
can be completely understood by decomposing
it into three independent point-lens events.  We begin by adopting this approach
not only for its didactic value but also because of its important implications
for the question of free-floating planets \citep{sumi11}.

The flux changes $F(t)$ in point-source/point-lens microlensing are described by five parameters,
$(t_0,u_0,t_{\rm E},f_s,f_b)$,
\begin{equation}
F(t) = f_sA(t) + f_b;
\qquad
A(t) = {u^2+2\over u\sqrt{u^2+4}};
\qquad
u(t) = \sqrt{\left({t-t_0\over t_{\rm E}}\right)^2 + u_0^2}.\label{equ:std}
\end{equation}
These are respectively the time of the closest approach, the impact
parameter in units of the Einstein radius $\theta_{\rm E}$, the
Einstein radius crossing time, the source flux, and the unmagnified
(or blended) flux within the aperture.

Lightcurves of the form of Equation~(\ref{equ:std}) fully describe both stellar
subevents.  However, the planetary lens transits the source, meaning
that the point-source formula (\ref{equ:std}) must be convolved with the source
profile in that case.  This introduces an additional parameter
\begin{equation}
\rho \equiv {\theta_*\over \theta_{\rm E}}
\end{equation}
where $\theta_*$ is the angular radius of the source.
The parameters of a separate fit for each subevent are shown in Table~\ref{tab:resPacz}. 
Note that we have allowed $f_b$ as a free parameter for the
final subevent (due to the secondary star
in OGLE-IV data), but have imposed $f_b=0$ in the remaining two (OGLE-III data)
subevents.

Setting $f_b$ at zero is appropriate under the assumption that the source
is the same for all three subevents.  
In principle, one must
consider that one or both of the OGLE-III subevents results from
lensing of a faint source that is blended with the principal, very
bright source.  However, given that less than 10\% of the light
comes from sources other than the one that was lensed in the OGLE-IV
subevent, this implies that $f_b/f_s>10$ for either of the other
subevents (if it is not due to the same bright source).  Imposing
this constraint leads to unacceptably bad fits for both subevents.
Note that \citet{jaroszynski10} found an acceptable binary-source fit to two OGLE-III
subevents before the OGLE-IV subevent was discovered.

The Einstein timescales $t_{{\rm E},i}$ 
(index $i$ numbers consecutive subevents)  
are related to the masses $M_i$ by
\begin{equation}
t_{\rm E} ={\theta_{\rm E}\over \mu};
\qquad
\theta_{\rm E} = \sqrt{\kappa M \pi_{\rm rel}};
\qquad \kappa \equiv {4 G\over c^2\rm AU} = 8.1~{\rm \frac{mas}{M_{\odot}}},
\end{equation}
where $\mu$ and $\pi_{\rm rel}$ are the lens-source relative proper motion and parallax, respectively. 
Because the internal motions of the system are much smaller than
the system motion across the line of sight, we approximate $\mu$
as the same for all sub-events.  Then the mass ratios between lenses
generating subevents are $q_{i,j}\equiv M_i/M_j = (t_{{\rm E},i}/t_{{\rm E},j})^2$.
We then find ratios $q_{1,2} = 2.66 \times 10^{-4}$ and $q_{3,2}=0.22$.
Similarly, the separation in units of $\theta_{\rm E,2}$ (i.e., primary) 
can be estimated from $s_{i,j} \simeq |t_{0,i}-t_{0,j}|/t_{\rm E,j}$, yielding
$s_{1,2} \simeq 5$ and $s_{3,2}\simeq 17$.  Even $\theta_{\rm E,2}$ can be estimated
once $\theta_*= 14.25\,\mu$as has been evaluated (see below) from
$\theta_{\rm E,2} = (\theta_*/\rho_1)(t_{\rm E,2}/t_{\rm E,1})= 0.33\,$mas.
This then gives an angular scale to the system and the basis to
estimate the lens masses and distance (see Section 4).  
The planet-host separation is $s_{1,2}\theta_{\rm E,2}\sim 1.6\,$mas, which
for lenses lying in the Galactic bulge corresponds to about $15~{\rm AU}$ (projected). 
In the same manner, the projected separation of the primary and the secondary 
stars is $48~{\rm AU}$. 
The assumption of bulge lens (as opposed to disk lens) gives also the primary mass of about $0.7~{\rm M_{\odot}}$, 
with an upper limit $M < 1.2~{\rm M_{\odot}}$, regardless of lens location.

The geometry of this system is illustrated in Figure~\ref{fig:traj}.
In the interest of maximum precision, we carry out more rigorous
calculations in the next section, but these confirm the very simple
arguments outlined here. 

\begin{deluxetable}{llrrr}
\tablewidth{0pt}
\tablecaption{Single Lens Parameters\label{tab:resPacz}}
\tablehead{
\colhead{quantity} &   
\colhead{unit} & 
\colhead{planetary} &
\colhead{primary} &
\colhead{secondary} \\
\colhead{} &   
\colhead{} & 
\colhead{subevent\tablenotemark{a}} &
\colhead{subevent\tablenotemark{b}} &
\colhead{subevent}
} 
\startdata
$t_0$ & d &       $4541.208 (29)$ & $4727.39(47)$ &     $5379.571(46)$ \\
$u_0$ &  &        $1.51 (20)$ &     $1.5502(65)$ &      $0.523(25)$ \\
$t_{\rm E}$ & d & $0.629 (58)$ &    $38.56(49)$ &       $17.94(52)$ \\
$\rho$ & &        $2.66 (13)$ &     $0$\tablenotemark{c} & $0$\tablenotemark{c}  \\
$f_b/f_s$ &  &    $0$\tablenotemark{d} &    $0$\tablenotemark{d}      & $0.016(73)$ \\ 
$\chi^2/dof$ & &  $372.0/349$ &     $368.6/343$ &       $427.2/378$ \\
\enddata
\tablenotetext{a}{The fit is based on OGLE-III data with primary lensing signal subtracted.}
\tablenotetext{b}{Seven OGLE-III epochs closest to planetary anomaly were removed before this fit was performed.}
\tablenotetext{c}{Data do not constrain value of $\rho$.}
\tablenotetext{d}{Value fixed based on secondary subevent results.}
\end{deluxetable}

\subsection{Detailed model \label{sec:det}} 

\begin{figure}
\includegraphics[bb=23 163 560 388,width=.98\textwidth]{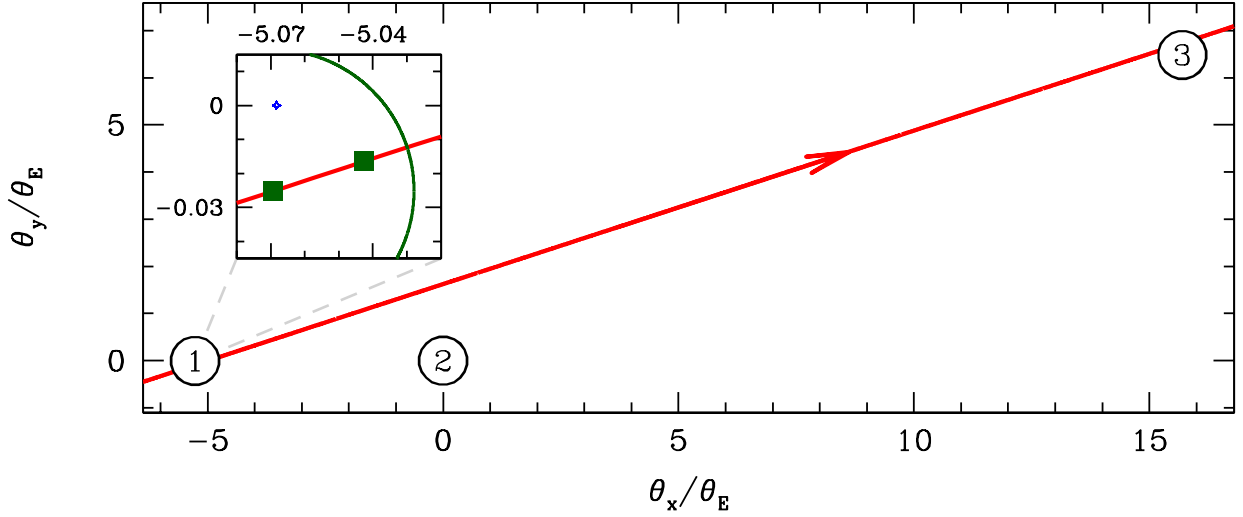}
\caption{Source trajectory relative to lens system 
(based on parameters derived in Section~\ref{sec:det}). 
Big circles show the positions of the primary (2; in the origin of the coordinate system),
the secondary (3), and the planet (1; located at $(-5.26,0)$).
Red line presents the trajectory and the direction of the source motion.
We assume here that the source passed the secondary caustic on the same side as the primary caustic,
but the data do not constrain this.
The axes are projected coordinates in units of $\theta_{\rm E}$ of the primary and both axes have the same scale.
The diamond-shaped planetary caustic
is presented in the inset according to scale (blue color;
other caustics are very close to the primary and the secondary and so were not approached by the source).
Source positions corresponding to the two epochs with the highest magnification are represented by green squares.
The green arc represents the source edge at the earlier of these two epochs.
\label{fig:traj}}
\end{figure}

Binary lenses have three parameters beyond those of the single lenses:
projected separation ($s_{1,2}$), 
mass ratio ($q_{1,2}$), 
and $\alpha_{1,2}$ -- the angle of the lens--source relative motion with respect to the binary axis. 
One can find more accurate than above (but still approximate) values of these parameters and $\rho$ by using simple geometry, 
scaling to appropriate $\theta_{\rm E}$, 
and correcting $s_{1,2} \rightarrow s_{1,2}-1/s_{1,2}$ \citep{han06}. 
These are compared to the results of the final modeling in Table~\ref{tab:resO3}. 
One finds very good agreement especially in the case of $s_{1,2}$. 
For final solution, the magnification during the planetary subevent was evaluated using 'mapmaking' method \citep{dong06} with modifications \citep{dong09b,poleski14a}. 
The magnification for other epochs was calculated by hexadecapole approximation \citep{pejcha09,gould08}. 
Note that OGLE-2008-BLG-092 is the first microlensing event for which different photometric datasets have no overlap in time. 
Such an overlap typically constrains the blending ratios for different datasets. 
Instead, we set a prior distribution of $f_b/f_s$ (implemented as a $\chi^2$ penalty) to be Gaussian with 0 mean and dispersion equal to the uncertainty of $f_b/f_s$ in the third subevent (Table~\ref{tab:resPacz}). 
Based on \citet{claret00}, we assumed the source limb darkening parameter $\Gamma_I = 0.546$ (or $u_I=0.643$). 
We checked that even a $10\%$ change of this parameter has a negligible impact on fitted values. 
The model parameters indicate that the planetary caustic was very small compared to the size of the source that passed over it. 
The source trajectory relative to planetary caustic is illustrated in the 
inset in  Figure~\ref{fig:traj}.

\begin{deluxetable}{llrrr}
\tablewidth{0pt}
\tablecaption{Binary Lens Parameters for OGLE-III Data (i.e., planet and host system)\label{tab:resO3}}
\tablehead{
\colhead{quantity} &   
\colhead{unit} & 
\colhead{value} &
\colhead{uncertainty} &
\colhead{approximated\tablenotemark{a}}
} 
\startdata
$t_0$ & d & 4727.374 & 0.047 & \\
$u_0$ &  & 1.545 & 0.043 & \\
$t_{\rm E}$ & d & 38.64 & 0.88 & \\
$\rho$ & & 0.0414 & 0.0025 & 0.0434 \\
$\alpha_{1,2}$ & deg & 17.99 & 0.30 & 18.06 \\
$s_{1,2}$ &  & 5.26 & 0.11 & 5.27 \\
$q_{1,2}$ & $10^{-4}$ & 2.41 & 0.45  & 2.66\\
$f_b/f_s$\tablenotemark{b} &  & 0.011 & 0.071 & \\
$\chi^2/dof$ & & $371.1/346$ & & \\ 
\enddata
\tablenotetext{a}{These values were estimated based on single lens fits to the three subevents (see Section~\ref{sec:det}).}
\tablenotetext{b}{Assumes $f_b/f_s = 0.000 \pm 0.073$ (see Section~\ref{sec:det}).}
\end{deluxetable}

Our best-fitting model predicts that the source passed the central and planetary caustics on opposite sides, but this is weakly constrained. 
The alternative model with the source passing both caustics on the same side has $\chi^2$ larger by $1.5$ and the only parameter that differs significantly is $\alpha_{1,2} = 17\fdg55(29)$, i.e., smaller by $0\fdg44$. 
We attempted to fit the model with the source passing relatively far away from the planetary caustic, i.e., the planetary subevent would be an almost standard 
point-source/point-lens 
light curve. 
In such cases, the value of $\rho$ would be smaller leading to very large value of $\mu$. 
However, this solution is ruled out ($\Delta\chi^2 = 76$). 
It is also consistent with the source star profile in the OGLE-I and OGLE-II reference images, which is circular. 

Instead of full triple lens modeling that is numerically complicated, we combine the results of the double lens model for the OGLE-III data (Table~\ref{tab:resO3}) 
and the single lens model for the OGLE-IV data (Table~\ref{tab:resPacz}) 
to obtain parameters of the triple lens. 
There are three additional parameters that must be estimated:
$q_{3,2}$ -- secondary to primary mass ratio, 
$s_{3,2}$ -- primary-secondary projected separation (in units of primary $\theta_{\rm E}$), and
$\phi_{2,3}$ -- the angle between the secondary and the planet as viewed from the primary. 
We checked that at $s_{3,2}\approx17$ the perturbations caused by the presence of one of the stars on the microlensing signal of the other one are much smaller than the photometric errors. 
The Einstein timescale is proportional to the square root of the lens mass, thus, 
$q_{3,2}$ is the squared ratio of primary and secondary $t_{\rm E}$ and equals $0.216(12)$. 
The values of $s_{3,2}$ and $\phi_{2,3}$ are calculated using simple geometry and noting that the caustics of both stars are $s_{3,2}-1/s_{3,2}$ apart. 
The source could have passed the stellar caustics on the same or opposite sides but the data do not constrain this. 
If they were passed on the same side (as illustrated in Figure~\ref{fig:traj}), we obtain $s_{3,2} = 16.99(38)$ and $\phi_{2,3} = 157\fdg59(35)$. 
The opposite situation leads to very similar value of $s_{3,2} = 17.03(38)$ and slightly smaller $\phi_{2,3} = 155\fdg95(36)$. 

In order to investigate the self-consistency of the triple lens model, we checked how the presence of the secondary affects the position of the planetary caustic that is the only caustic that was approached closely. 
We fixed the mass ratios as well as positions of the primary and the secondary and tried 
to find the position of the planet that gives the same position and size of the planetary caustic, as we have in the double lens model. 
It was enough to change the planet position by as little as $0.01$ to match the caustics. 
This is much less than $0.11$  uncertainty of $s_{1,2}$. 
Therefore, we conclude that our approach to the triple lens modeling is valid and available data do not require more complicated methods.

The models presented above neglected the microlensing parallax ($\pi_{{\rm E}}$) effect. 
We tried to incorporate $\pi_{\rm E}$ in the single lens models fitted to the OGLE-IV data but the short event timescale renders the parallax effect too weak to measure. 
The $\Delta\chi^2 = 9$ limit is $\pi_{{\rm E}} < 3.6$ for negative $u_{0}$ and $\pi_{{\rm E}} < 3.1$ for positive $u_{0}$. 
The low magnification of the primary event in the OGLE-III data also does not allow $\pi_{\rm E}$ to be measured. 
For similar reasons, the effect of the lens rotation was ignored. 

There is one more possibility  that one should consider. 
While three separate subevents were observed and these had to be caused by three different lenses, this does not necessarily imply that these three objects are bound. 
The secondary subevent could have been caused by a star that is unrelated to the planet-host system. 
Such an unrelated lens can give rise to microlensing of a given source with probability similar to the general event rate of $10^{-5}$ per year. 
For the secondary subevent happening within two years before or after the main event, the probability is about $4\times10^{-5}$. 
On the other hand, 
the frequency of binary systems with separations equal or smaller than the deprojected separation of OGLE-2008-BLG-092L is on order of $40\%$ \citep{duquennoy91}. 
However, the putative companion gives rise to the microlensing event only if the source trajectory passes close enough i.e., $u_{0,3} < 1$.  
In the case of OGLE-2008-BLG-092L the source had to enter or exit (factor of 2) the circle of circumference $2\pi s_{3,2}\theta_{\rm E,2}$ within an arc of length $2\theta_{\rm E,3}$. 
Thus, the probability that the secondary is detected is $4\theta_{\rm E,3}/\left(2\pi s_{3,2}\theta_{\rm E,2}\right) = 2t_{\rm E,3}/(\pi s_{3,2}t_{\rm E,2}) \approx 0.02$. 
In this sample of binary system most have smaller separations than considered above, which leads to higher probability of detecting the secondary, which we assume here to be $0.04$.
Finally, $0.4\times0.04 = 0.016$ is 400 times larger than $4\times10^{-5}$, 
which makes unbound lens scenario highly unlikely. 

\section{Physical properties} 

\subsection{Source star properties} 

\begin{figure}
\epsscale{.79}
\plotone{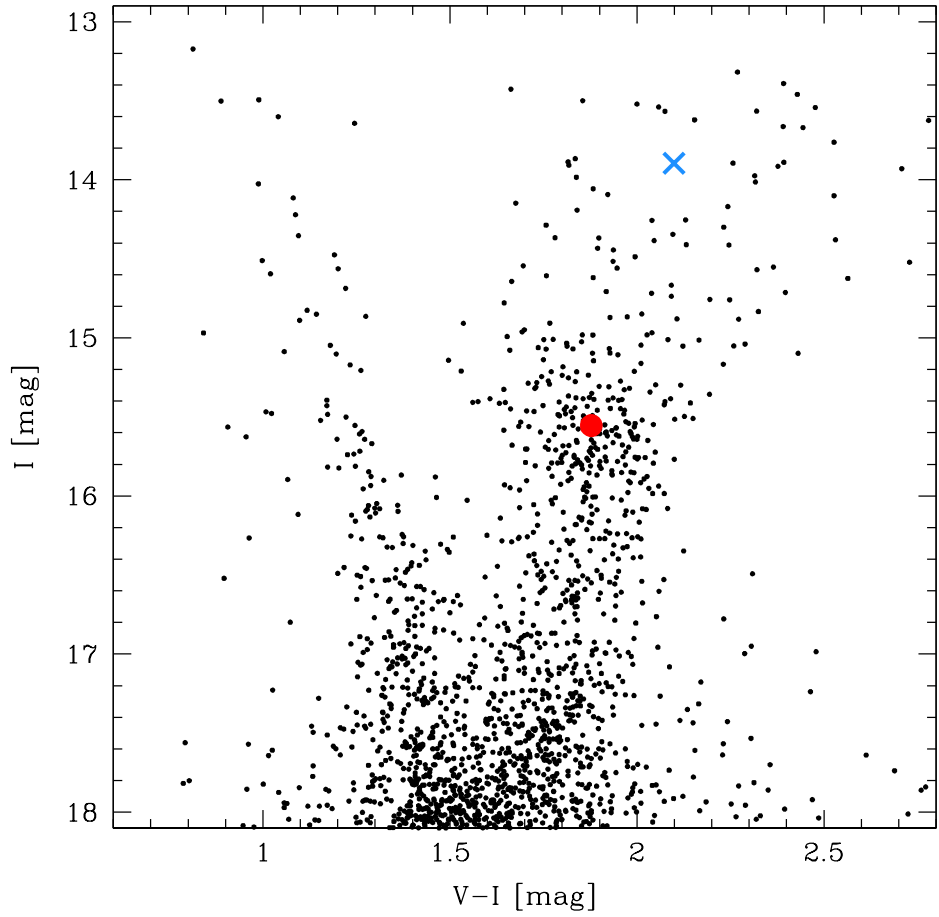}
\caption{Color-magnitude diagram for stars within $1\farcm7$ of the event. 
Red circle marks the red clump centroid while blue cross represents the unmagnified source position. 
\label{fig:cmd}}
\end{figure}

The first step required to translate microlensing parameters to physical parameters is estimating the angular source radius ($\theta_{\star}$). 
The source has observed $I$-band brightness of $13.90~{\rm mag}$ and $(V-I)$ color of $1.88~{\rm mag}$ and almost all stars at this part of the color-magnitude diagram (Figure~\ref{fig:cmd}) are located in the bulge.  
The red clump intrinsic $(V-I)$ color is $1.06~{\rm mag}$ \citep{bensby11} and $l = -4\fdg75$ implies dereddened $I$-band brightness of $14.61~{\rm mag}$ \citep{nataf13b}. 
This gives source reddening corrected $V$-band brightness of $14.23~{\rm mag}$ and $(V-I)$ color of $1.28~{\rm mag}$. 
The corresponding $(V-K)$ color is $2.88~{\rm mag}$ \citep{bessell88}. 
The color-surface brightness relation \citep{kervella04b} leads to $\theta_{\star} = 14.25(60)~{\rm \mu as}$. 
The Einstein ring radius is 
\begin{equation}
\theta_{\rm E} = \theta_{\star}/\rho = 0.344(20)~{\rm mas}.
\end{equation}
This results in 
\begin{equation}
\mu = \theta_{\rm E}/t_{\rm E} = 3.25(22)~{\rm mas/yr}.
\end{equation}

The baseline flux comes almost entirely from the source. Thus, we can use eight years of the OGLE-III monitoring to measure the source proper motion. 
Its value relative to the local red clump stars is 
$\mu_{\star,\alpha\cos\delta} = 0.25(31)~{\rm mas/yr}$ and 
$\mu_{\star,\delta} = 1.17(31)~{\rm mas/yr}$. 
This corresponds to 
$\mu_{\star,l\cos b} = 1.13(31)~{\rm mas/yr}$ and 
$\mu_{\star,b} = 0.39(31)~{\rm mas/yr}$ in Galactic coordinates. 

\subsection{Lens system properties} 

The lens parallax and mass are \citep{gould00b}:
\begin{equation}
\pi_l = \pi_{rel} + \pi_s,
\qquad \pi_{rel} = \pi_{\rm E}\theta_{\rm E}
\end{equation}
\begin{equation}
M = \frac{\theta_{\rm E}}{\kappa\pi_{\rm E}},
\end{equation}
where $M$ is the primary mass (because $\theta_{\rm E}$ corresponds to primary lens).  
Both the source position in the color-magnitude diagram and proper motion are consistent with a bulge location. 
For some previously published planets the value of $\pi_{\rm E}$ was found from the 
microlensing model fit. 
Despite three subevents observed,  one of them having an Einstein timescale longer than a month, we were not able to make a meaningful measurement of $\pi_{\rm E}$. 
The limits on $\pi_{\rm E}$ from OGLE-IV data translate to uninteresting limits on lens mass $M > 0.012~{\rm M_{\odot}}$ and distance $D_l > 0.75~{\rm kpc}$. 

Instead of finding $\pi_{\rm E}$ and calculating $D_l$ and $M$ from this, we constrain $D_l$ and $D_s$ 
using the indirect evidence based on $\theta_{\rm E}$ and $\mu$. 
The value of $\mu = 3.2~{\rm mas/yr}$ is comparable to the dispersion of the proper motions in the Galactic bulge in one direction \citep{vieira07}. 
This strongly suggests that the source and the lens are relatively close to each other, i.e., the lens lies in the bulge or close to its edge. 
We do not know the orientation of $\mu$ relative to the source proper motion $\mu_{\star} = 1.2~{\rm mas/yr}$ but even in the most unfavorable case, the bulge lens hypothesis is preferred over a disk lens, 
because the typical proper motions of disk stars are $6.5~{\rm mas/yr}$. 

\begin{figure}
\epsscale{.79}
\plotone{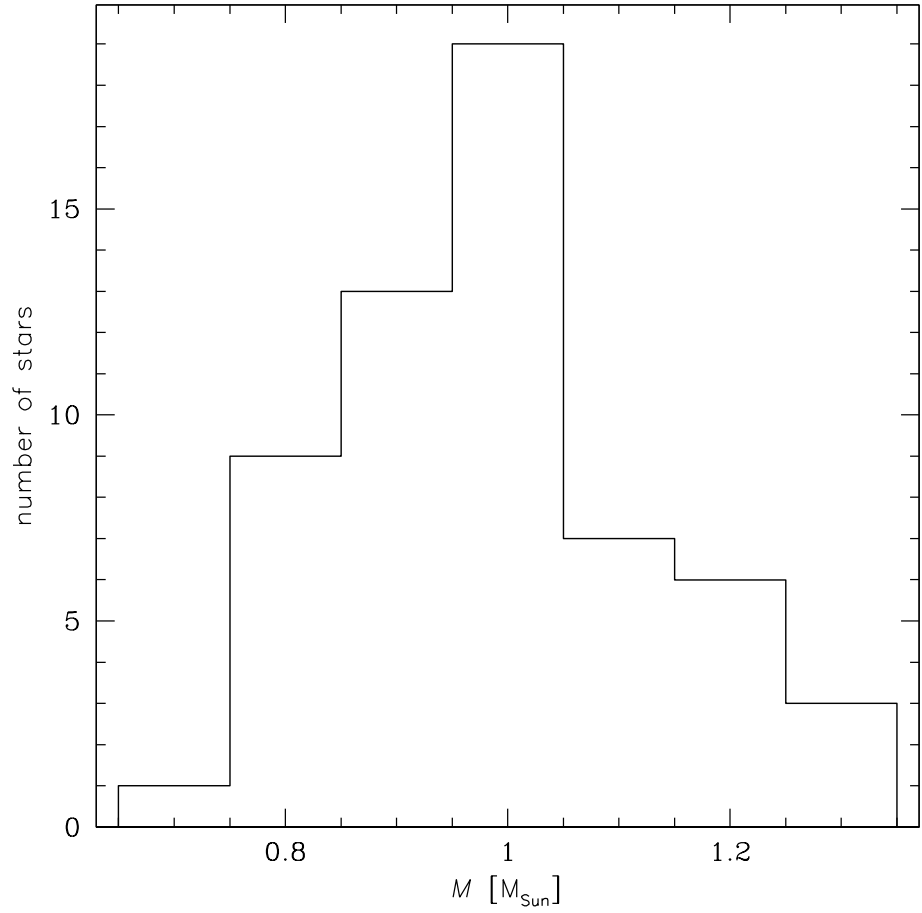}
\caption{Histogram of masses of the bulge MS and subgiant stars observed by \citet{bensby13}. 
\label{fig:bensby}}
\end{figure}

Noting that the lens is in the bulge we can put the upper limit on the lens mass. 
The problem of the bulge highest mass main main sequence (MS) stars was not investigated previously from observers' point of view. 
There are no direct mass measurements for bulge MS stars. 
Masses can be estimated based on the high-resolution and high signal-to-noise spectra, but these require extremely long observations even at the largest existing telescopes for bulge MS targets. 
This can be overcome by observing the high-magnification microlensing events with the MS sources close to the peak. 
\citet{bensby13} presented the most recent analysis of spectra from the ongoing project aimed at systematic spectroscopic observations of the microlensing events with subgiant or MS sources. 
In Figure~\ref{fig:bensby} we present the histogram of derived masses. 
We see that most of the values are close to $1~{\rm M_{\odot}}$ and the highest measured masses are around $1.3~{\rm M_{\odot}}$. 
This distribution is significantly affected by the selection bias, as lower mass stars are too faint to be observed. 
Taking into account measurement errors in spectroscopically derived masses and the fact that these can be both subgiant and MS stars, we estimate the bulge MS stars have masses below $1.2~{\rm M_{\odot}}$ or probably even below $1.1~{\rm M_{\odot}}$.  
The former limit corresponds to a planet mass $q_{1,2}M$ close to that of Saturn. 

The distances to the source and the lens were estimated using a simulation of microlensing events. 
Only the events with both sources and lenses in the bulge were simulated. 
The density of stars was modeled according to \citet{nataf13b} results at the vicinity of the event, i.e., with mean distance of $8.86~{\rm kpc}$ and Gaussian distribution of distance modulus with a dispersion of $0.263~{\rm mag}$. 
The probability of observing microlensing event depends not only on the density of stars but also on the size of the $\theta_{\rm E}$, thus, the simulated events were additionally weighted by $\sqrt{1/D_l-1/D_s}$. 
We rejected the events that resulted in lens masses above $1.2~{\rm M_{\odot}}$. 
Integrating the lens and source distances in the rest of the simulated events leads to $D_l=8.1~{\rm kpc}$ and $D_s=9.7~{\rm kpc}$. 

Then the Einstein radius in physical units is $D_l\theta_{\rm E} = 2.8~{\rm AU}$. 
The projected separation of the planet is $15~{\rm AU}$, 
while the secondary is at the projected separation of $48~{\rm AU}$. 
These values can be de-projected by multiplying by $\sqrt{3/2}$, which yields $18~{\rm AU}$ and $58~{\rm AU}$, respectively. 
The adopted distances $(D_l, D_s) = (8.1, 9.7)~{\rm kpc}$ yield $\pi_{rel} = 20~{\rm mas}$ and therefore 
$M=\theta_{\rm E}^2/\kappa\pi_{rel} = 0.71~{\rm M_{\odot}}$. 
This gives the planetary mass of $q_{1,2}M = 3.3~{\rm M_{Neptune}} = 3.9~{\rm M_{Uranus}}$. 
The mass of the secondary is $q_{3,2}M = 0.15~{\rm M_{\odot}}$. 


\section{Discussion} 

\subsection{Planet properties and formation}

The ratio of the de-projected separation 
to the snow-line distance is
\begin{equation}
\frac{\sqrt{3/2}s_{1,2}\theta_{\rm E}D_l}{2.7\left(\frac{M}{\rm M_{\odot}}\right){\rm AU}} = 56 \times \left(1-\frac{D_l}{D_s}\right).
\end{equation}
For the inferred distances $D_l=8.1~{\rm kpc}$ and $D_s=9.7~{\rm kpc}$ this ratio is about nine, making OGLE-2008-BLG-092LAb the first extrasolar analog of Uranus. 
The above ratio would be closer to values found in solar system gas giants as opposed to ice giants if $D_l/D_s > 0.91$. 
This condition is almost equivalent to already excluded values of $M>1.2~{\rm M_{\odot}}$.
The simulation of bulge-bulge events presented above indicates the probability of $D_l/D_s > 0.91$ is marginal. 

Uranus and Neptune were formed closer to the Sun and migrated outward because of the interaction with Jupiter and Saturn. 
The key factors that led to this conclusion were planetary masses and separations. 
If the properties of OGLE-2008-BLG-092LAb are taken at face value and combined with a host mass that is smaller than the Sun, they suggest in situ planet formation is not possible. 
However, the density of the protoplanetary nebula of OGLE-2008-BLG-092LA could be higher than the solar one. 
This would lead not only to formation of OGLE-2008-BLG-092LAb, but also other massive planets in this system. 
The sensitivity for detecting such additional planets was very low, thus we cannot rule out the hypothesis of much denser protoplanetary nebula.

The planet could be also formed closer to the host and migrated outward. 
In hierarchical triple system the Lidov-Kozai effect \citep{lidov62,fabrycky07} changes the eccentricity of the inner orbit in oscillatory way. 
For a short period of time, the eccentricity of the inner binary is very large resulting in short pericenter distance. 
During the pericenter passage the tidal forces can only shrink the orbit, not expand it. 
Alternatively, 
the system may contain more planets, whose role might then have been similar to Jupiter and Saturn in the solar system. 
It is also possible that the planetary orbit has significant eccentricity. 
This would affect the relation between the semi-major  axis and observed projected separation, leading to semi-major axis that is overestimated. 
With more examples of similar planets observed in the future, more definitive answer to the question of their origin could be derived.

\subsection{Prospects for detecting similar planets}

Detecting  cold ice giants that, like Uranus and Neptune are many AU away from their hosts, is  extremely difficult for techniques other than microlensing. 
In the case of the radial velocity method, a very long survey would be required to accomplish this. 
Currently, the longest period transiting planet is Kepler-421b with a period of nearly two years \citep{kipping14}. 
This period is still about $40$ times shorter than required to find a Uranus analog. 
There are a few candidates for longer orbital periods found by the {\sl Kepler} satellite \citep{burke14}. 
For some stars, even the longest observing campaigns conducted, reveal only one transit. 
Although the orbital period cannot be estimated based on photometry alone, 
combining transit timing with 
radial velocity measurements may  constrain the orbital period \citep{yee08}. 
Such a combination of transit and radial velocity methods 
is still limited by the low probability that a planet on a very wide orbit transits in front of the host star. 
Ice giants orbiting nearby stars can have angular separations larger than the resolution achieved by direct imaging. 
However, 
the problem lies in detecting the light from the planet, which cannot be done by any of the instruments that is currently working or planned for the near future. 

We showed above that microlensing is capable of discovering planets on very wide orbits. 
There are a few different channels that can lead to microlensing planet discoveries. 
First, the source may approach very close to the lens. 
This gives rise to very large magnification and the event is sensitive to even small perturbations in the magnification pattern caused by the presence of the lens companions \citep{griest98}. 
Such close proximity can be predicted while the event is ongoing, and intensified observations lead to greater sensitivity for planets \citep[e.g.,][]{udalski05,gould10}. 
This channel led to most of the planets published to date \citep{zhu14}, but its sensitivity drops for planets with large $s$. 
The size of central caustic is $w = \frac{4q}{\left(s-s^{-1}\right)^2}$ \citep{chung05} and 
for OGLE-2008-BLG-092, it would result in $w = 3.8 \times 10^{-5}$ if the secondary star were not present. 
Such small caustics cannot be detected by current microlensing experiments, 
because the signal is smeared by the much larger source \citep{chung05}. 
Second, the planet can be revealed if the source approaches a resonant caustic, which  is large and forms if  $s\approx1$. 
This requires  that the Einstein ring has a size larger than $10~{\rm AU}$, which is highly unlikely. 
Third, the source may approach the planetary caustic, that is located close to the planet \citep{mao91,gould92,distefano99}. 
Such approaches are unpredictable and well separated in time from the host lensing event. 
Follow-up photometry can be obtained after the anomaly is detected or by routine monitoring of many events after the peak, which is not done currently. 
This shows that the microlensing surveys that cover significant part of the bulge with high cadence are the only efficient way to investigate the population of extrasolar ice giants. 

There are further obstacles to  detecting microlensing planets at large separations. 
The typical microlensing event has $t_{\rm E} \approx 25~{\rm d}$,  
which means that a significant number of planetary caustic approaches are a few months before or after the main peak. 
They may be unobservable because of the conjunction of the Sun and the bulge. 
They may also happen when the Sun is relatively close to the bulge so that $24$ hour a day observations are not possible. 
We showed that even a few years ago, the OGLE survey had cadence good enough to identify such an anomaly. 
At present, the OGLE-IV survey has sufficient cadence to observe events similar to OGLE-2008-BLG-092 in $26~{\rm deg}^2$ of bulge. 
The survey observing strategy is well defined, therefore, a larger sample of similar planetary caustic approaches would allow one to estimate the frequency of such planets.

\subsection{Planetary anomaly and free-floating planets}

One of the important outcomes of the microlensing experiments was a detection of an excess of the events with very short $t_{\rm E}$ of $1 - 2$ days \citep{sumi11}. 
Since $t_{\rm E}$ is proportional to the square root of the lens mass, these events should be produced by planetary mass objects. 
\citet{sumi11} presented 10 events with very short $t_{\rm E}$ and no signs of other lenses, and therefore, interpreted them as free-floating planets with abundance of $1.8^{+1.7}_{-0.8}$ per MS star. 
There are other possible interpretations of these events. 
Some of them may involve bound planets on very wide orbits, for which the presence of the host was not revealed based on either additional bump in photometry \citep{han05a} or distortion of planetary event \citep{ryu13}.
Based on the direct imaging upper limits on the abundance of Jupiter-mass planets on $10-250~{\rm AU}$ wide orbits \citep{lafreniere07}, \citet{sumi11} estimated that no more than $25\%$ of detected short events are caused by bound planets. 
However, we find a significant observational bias in detecting the stellar hosts for these planets. 
Even if the microlensing planet is bound, the signal from the host may be too weak to be measured. 
In the case of OGLE-2008-BLG-092, even a $1\%$ signal would be revealed because the source is very bright. 
This corresponds to $u_{0,2}<3.5$ for host lensing. 
For source trajectories crossing the planetary caustic with randomly selected angle relative to the planet-host axis, the probability that the source would make a closer approach to the host is $0.49$. 
For $10\%$ magnification ($u_{0,2}<1.7$) the corresponding probability is $0.21$. 
These numbers ignore the lack of data due to seasonal gaps. 
It is harder to detect host subevent if the photometric accuracy is lower. 
The source stars in \citet{sumi11} sample were at least $3.3~{\rm mag}$ fainter than OGLE-2008-BLG-092S and $1\%$ signals could be easily missed. 

We note that, the events similar to OGLE-2008-BLG-092 but with fainter sources (i.e., smaller $\rho$) and different $\alpha_{1,2}$, would not pass the selection criteria used by \citet{sumi11}. 
One of their requirements was $u_0<1$ and we measured $u_{0,1} = 1.51(20)$ in single lens approximation. 

\citet{sumi11} presented also $2\sigma$ detection limits for hosts of each presented planet based on the light curve distortions.  
Only two out of ten cases are inconsistent with
a planet that is similar to OGLE-2008-BLG-092LAb i.e., with $s=5.26$. 
These are MOA-ip-1 and MOA-ip-10. 
The third case (MOA-ip-7) is marginally consistent ($s>5.2$). 
We note that  
there are two other events that showed separate subevents by planetary microlensing. 
First, \citet{skowron09} analyzed the event OGLE-2002-BLG-045. 
The best fitting model was a binary lens with $s=4.0$ and $q=0.008$. 
However, the analyzed data were very limited. 
Even though the mass ratio was found to be in planetary regime, this event is not considered a secure planet detection. 
Second, \citet{bennett12} presented event MOA-bin-1 
that is best fitted by a planet with $s=2.10$ and $q=0.0049$. 
In this case, the host lensing signal was not detected, but the evidence for its presence is a distortion of the planetary anomaly.

\subsection{Long-term stability of the system}

We have discovered a relatively compact triple system and thus its long-term dynamical stability should be investigated. 
\citet{holman99} investigated the conditions under which a massless planet 
orbiting one component of a stellar binary is a stable configuration. 
The initial planetary orbit was assumed to be circular, prograde, and coplanar with the stellar binary.
A wide range of stellar binary eccentricities ($e$) was tested. 
Based on results of \citet{holman99} and $q_{3,2}$ in OGLE-2008-BLG-092L 
we obtain the following stability criterion
($a_c$ -- critical planet semi-major axis, $a_b$ -- stellar binary semi-major axis): 
\begin{equation}
 \frac{a_c}{a_b} = 0.397(7)-0.527(36) e + 0.115(43)e^2
\end{equation}
where the quoted uncertainties take into account uncertainty of $q_{3,2}$ as well as those from the formula in \citet{holman99}. 
We have measured $s_{1,2}/s_{3,2} = 0.3096(95)$, 
thus the system is stable if the measured projected separation ratio represents the semi-major axis ratio 
and the stellar orbit is circular. 
However, we must note that $s_{1,2}$ and $s_{3,2}$ correspond to two different epochs. 
The planetary orbit remains stable for binary eccentricities up to $0.17$. 
If the stellar binary is at its apoastron then planetary orbit is stable for $e < 0.11$. 

\section{Summary} 

The microlensing event OGLE-2008-BLG-092 showed three separate subevents that reveal a wide stellar binary system with a planet orbiting the primary star. 
The planet turns out to be the first extrasolar analog of Uranus i.e., a cold ice giant that orbits the parent star at a distance many times larger than the snow-line distance. 
Thus, the planet likely also shares a similar origin with solar System ice giants. 
This discovery opens a new part of the planet properties parameter space. 
We also showed that no method other than microlensing can be used to find such planets. 
Among different scenarios that lead to microlensing planet detection, only the approach to the planetary caustic gives a  high chance for detecting planets on wide orbits. 
The contribution of bound planets to the very short events sample should be carefully investigated in the future. 
The detection of the  bump due to a putative host can be hindered by gaps in the data and lower photometric accuracy for fainter sources. 
These should be considered together with constraints from the anomaly distortions to find unbiased abundances of both free-floating and bound planets.

\acknowledgments

We thank Scott Gaudi, Todd Thompson, Jennifer Yee, and Wei Zhu for consultation 
and anonymous referee for useful suggestions. 
We acknowledge funding from the European
Research Council under the European Community’s Seventh
Framework Programme (FP7/2007-2013)/ERC grant agreement no. 246678 to A.U. for OGLE project. 
R.P. acknowledges support by The Thomas Jefferson Chair for Discovery and Space Exploration. 
This research was partly supported by the Polish Ministry of 
Science and Higher Education (MNiSW) through the program ``Iuventus Plus'' award no. IP2011 043571 and  IP2011 026771. 
The work by C.H. was supported by the Creative Research 
Initiative Program (2009-0081561) of National Research Foundation of Korea.
S.D. was supported by ``the Strategic Priority Research Program-The Emergence 
of Cosmological Structures'' of the Chinese Academy of Sciences (Grant No. XDB09000000). 
A.G. acknowledges support by NSF grant AST 1103471.


\bibliographystyle{apj}

\end{document}